\documentstyle[prl,aps,multicol,epsf]{revtex}
\begin {document}
\title{Fractional Josephson vortices at YBa$_2$Cu$_3$O$_{7-x}$ grain
boundaries}
\author{R.~G.~Mints and Ilya~Papiashvili}
\address{School of Physics and Astronomy,\\
Raymond and Beverly Sackler Faculty of Exact Sciences, Tel Aviv
University, Tel Aviv 69978, Israel}
\date{\today}
\maketitle
\begin{abstract}
We report numerical simulations of magnetic flux patterns in asymmetric
45$^{\circ}$ [001]-tilt grain boundaries in YBa$_2$Cu$_3$O$_{7-x}$
superconducting films. The grain boundaries are treated as Josephson
junctions with the critical current density $j_c(x)$ alternating along
the junctions. We demonstrate the existence of Josephson vortices with
fractional flux quanta for both periodic and random $j_c(x)$. A method
is proposed to extract fractional vortices from experimental flux
patterns.
\end{abstract}
\begin{multicols}{2}
\narrowtext
Numerous recent studies of electromagnetic properties of grain
boundaries in high-$T_c$ superconducting films are driven by
necessity to probe the fundamental symmetry of the order parameter
and the flux quantization\cite{Lin,Tsu,Mil}. Although
interpretation of the results is nontrivial, most of the data can
be understood in terms of the conventional model of a strongly
coupled Josephson junction\cite{Gro}. The asymmetric 45$^{\circ}$
[001]-tilt grain boundary in YBa$_2$Cu$_3$O$_{7-x}$ films is a
notable exception of this rule. We show in this letter that these
boundaries should have unusual electromagnetic properties
unprecedented for the physics of standard Josephson junctions.
\par
Experimentally the asymmetric 45$^{\circ}$ [001]-tilt grain boundaries
in YBa$_2$Cu$_3$O$_{7-x}$ films exhibit an anomalous dependence of the
critical current $I_c$ on the applied magnetic field $H_a$
\cite{Cop,Man}. Contrary to the usual Fraunhoffer-type dependence with
a major central peak at $H_a=0$ and minor symmetric side-peaks, these
boundaries demonstrate a pattern without the central major peak.
Instead, two symmetric major side-peaks appear at certain fields
$H_a=\pm\,H_p\ne 0$ \cite{Cop,Man,Min}. Other remarkable feature is
the spontaneous disordered magnetic flux generated at the asymmetric
45$^{\circ}$ [001]-tilt grain boundaries in YBa$_2$Cu$_3$O$_{7-x}$
films \cite{JMa}. It is worth noting that the spontaneous flux is
observed only in samples exhibiting the anomalous dependence
$I_c(H_a)$.
\par
Clearly, the major side-peaks reveal a {\it specific} heterogeneity of
the Josephson properties. Indeed, a fine scale faceting of grain
boundaries in YBCO thin films has been recorded by the transmission
electron microscopy \cite{Mil,Jia,Ros,Tra}. The facets have a typical
length-scale $l$ of the order of 10--100 nm and a wide variety of
orientations. This grain boundary structure combined with a
predominant $d_{x^2-y^2}$ wave symmetry of the order parameter
\cite{Tsu,Wol,Bra,Van} form a basis for understanding both the
anomalous dependence $I_c(H_a)$ and the spontaneous flux
\cite{Min,JMa,RGM,MP}.
\par
In the case of a $d_{x^2-y^2}$ wave superconductor the phase difference
of the order parameter across the grain boundary consists of two
terms. The first term $\varphi (x)$ is caused by a magnetic flux
inside the junction and the second term $\alpha (x)$ is caused by a
misalignment of the anisotropic banks of the junction. The Josephson
current density $j(x)$ depends on the total phase difference
$\varphi(x) +\alpha(x)$. Assuming $j(x)\propto\sin\,[\varphi (x)+\alpha
(x)]$ one can develop a model of the electromagnetic properties of the
grain boundaries in YBCO films \cite{Man}. Values of the phase $\alpha
(x)$ depend on the relative orientation of the neighboring facets. In
the case of an asymmetric faceted 45$^{\circ}$ grain boundary we have
an interchange of $\alpha =0$ and $\pi$ and $j(x)=j_c(x)\sin\varphi
(x)$, where the alternating critical current density
$j_c(x)\propto\cos\alpha(x)$.
\par
In this letter we report numerical simulations of flux patterns in the
asymmetric 45$^{\circ}$ [001]-tilt grain boundaries in
YBa$_2$Cu$_3$O$_{7-x}$ superconducting films. The boundaries are
treated as Josephson junctions with an alternating critical current
density $j_c(x)$. We find two types of fractional Josephson vortices
for each stationary state with a spontaneous flux in the grain
boundaries, which exists for both periodic and random sequences of
facets. One type of vortices contains the magnetic flux
$\phi_1<\phi_0/2$; the other type carries $\phi_2>\phi_0/2$ with a
complementarity condition $\phi_1 +\phi_2=\phi_0$, where $\phi_0$ is
the flux quantum. We suggest a method to extract the fractional
vortices from the data on flux patterns.
\par
The alternating dependence $j_c(x)$ is imposed by a particular sequence
of facets along the boundary and therefore $j_c(x)$ has the same
typical length-scale $l$ as the facets. It is convenient for the
further analysis to write $j_c(x)$ as
\begin{equation}
j_c=\langle j_c\rangle\,[1+g(x)],
\label{eq1}
\end{equation}
where $\langle j_c\rangle$ is the average value of the critical current
density over distances $L\gg l$:
\begin{equation}
\langle j_c\rangle ={1\over L}\,\int_0^Lj_c(x)\,dx.
\label{eq2}
\end{equation}
\par
The dimensionless function $g(x)$ in Eq.~(\ref{eq1}) characterizes the
Josephson properties of the grain boundary. This function alternates
with a typical length-scale $l$ and has a zero average: $\langle
g(x)\rangle =0$. The maximum amplitude of alternations of $|g(x)|$ may
vary from $|g(x)|_{\rm max}\gtrsim 1$ to $|g(x)|_{\rm max}\gg 1$
depending on details of the structure of the faceted grain boundary.
We assume that $\lambda\ll l\ll\Lambda_J$, where $\lambda$ is the
London penetration depth and
\begin{equation}
\Lambda_J^2={c\,\phi_o\over 16\pi^2\lambda\langle j_c\rangle}.
\label{eq3}
\end{equation}
is an effective Josephson penetration depth. With this notation, the
phase difference $\varphi(x)$ satisfies
\begin{equation}
\Lambda_J^2\,\varphi''-[1+g(x)]\sin\varphi=0.
\label{eq4}
\end{equation}
\par
A model grain boundary with a periodic critical current density
$j_c(x)$ has been considered analytically by means of a two-scale
perturbation theory which requires $l\ll\Lambda_J$\cite{RGM}. In this
approximation, the phase $\varphi(x)$ is a sum of a smooth part
$\psi(x)$ with a length scale $\Lambda_J$ and a rapidly oscillating
part $\xi(x)$ with a length scale $\l$ and a small amplitude
$|\xi(x)|\ll 1$:
\begin{equation}
\varphi(x)=\psi(x)+\xi(x).
\label{eq5}
\end{equation}
We have for the phases $\psi(x)$ and $\xi(x)$:
\begin{eqnarray}
\Lambda_J^2\psi ''-\sin\psi +\gamma \sin\psi\cos\psi=0,
\label{eq6}\\
\xi(x) =\xi_g(x)\sin\psi,
\label{eq7}
\end{eqnarray}
where the function $\xi_g(x)$ is given by
\begin{equation}
\Lambda_J^2\xi_g''=g(x),
\label{eq8}
\end{equation}
and the dimensionless parameter $\gamma >0$ is defined as
\begin{equation}
\gamma =-\langle g(x)\xi_g(x)\rangle
=\Lambda_J^2\langle{\xi'_g}^2\rangle .
\label{eq9}
\end{equation}
It is worth mentioning that both $\xi_g(x)$ and $\gamma$ depend only
on the spatial distribution of $j_c$ and therefore characterize the
{\it individual} Josephson properties of a particular grain boundary.
We stress that this approximation is valid if $l\ll\Lambda_J$ and
$|\xi(x)|\ll|\psi(x)|$ (the latter condition results in $|g(x)|\ll
4\pi^2\Lambda^2_J/l^2$).
\par
In the framework of the two-scale perturbation theory, a single
Josephson vortex is described by the solution of Eq.~(\ref{eq6}) under
the boundary conditions $\psi'(\pm\infty)=0$. The latter can be written
as $\sin\psi_\pm\,(1-\gamma\cos\psi_\pm)=0$, where
$\psi_\pm=\psi(\pm\infty)$. It is convenient for the further analysis
to assume that $\psi_-<\psi_+$.
\par
In the case of $\gamma <1$, there is only one single vortex solution,
for which the phase $\psi (x)$ increases monotonically from $\psi_-=0$
to $\psi_+=2\pi$. This solution describes the Josephson vortex with one
flux quantum $\phi_0$. In the case of $\gamma >1$, the spatial
distribution of the smooth phase $\psi$ describes two fractional
vortices. For the {\it first} fractional vortex the phase $\psi(x)$
increases from $\psi_-=-\psi_\gamma$ to $\psi_+=\psi_\gamma$, where
\begin{equation}
\psi_{\gamma} = \arccos(1/\gamma).
\label{eq10}
\end{equation}
The difference $\psi_+-\psi_-=2\psi_\gamma$ and thus this vortex
carries the flux $\phi_1=\psi_\gamma\phi_0/\pi<\phi_0/2$. For the {\it
second} fractional vortex $\psi_-=\psi_\gamma$, $\psi_+=2\pi -
\psi_\gamma$, the phase difference being $2\pi -2\psi_\gamma$, and thus
this vortex contains the flux
$\phi_2=(1-\psi_\gamma/\pi)\phi_0>\phi_0/2$. These two fractional
vortices are {\it complementary}: $\phi_1 +\phi_2 =\phi_0$.
\par
In our numerical study we solve Eq.~(\ref{eq4}) exactly. We treat
the stationary states as well as the relaxation to the stationary
states using a time-dependent model\cite{MP}
\begin{equation}
\ddot\varphi +\alpha\dot\varphi -\varphi'' +[1+g(x)]\sin\varphi=0,
\label{eq11}
\end{equation}
where $\alpha$ is a decay constant which we take from the interval
$0.1<\alpha <1$. The term $\alpha\dot\varphi$ in Eq.~(\ref{eq11})
describes dissipation driving the system into one of the stable
stationary states described by the solutions of Eq.~(\ref{eq4}).
\par
We begin our numerical simulations with verification of the results
obtained by means of the two-scale approximation for the grain boundary
with periodic $j_c(x)$. To study the fractional vortices we start the
simulations from a certain initial phase $\varphi_i(x)$ under the
condition $\varphi_i({\cal L})-\varphi_i(0) = 2\pi n$, where the
boundary length ${\cal L}\gg\Lambda_J$. In this case the numerical
procedure converges well to a final stationary state.
\par
\begin{figure}
\epsfxsize 0.95\hsize
\vskip .25\baselineskip
\centerline{\epsfbox {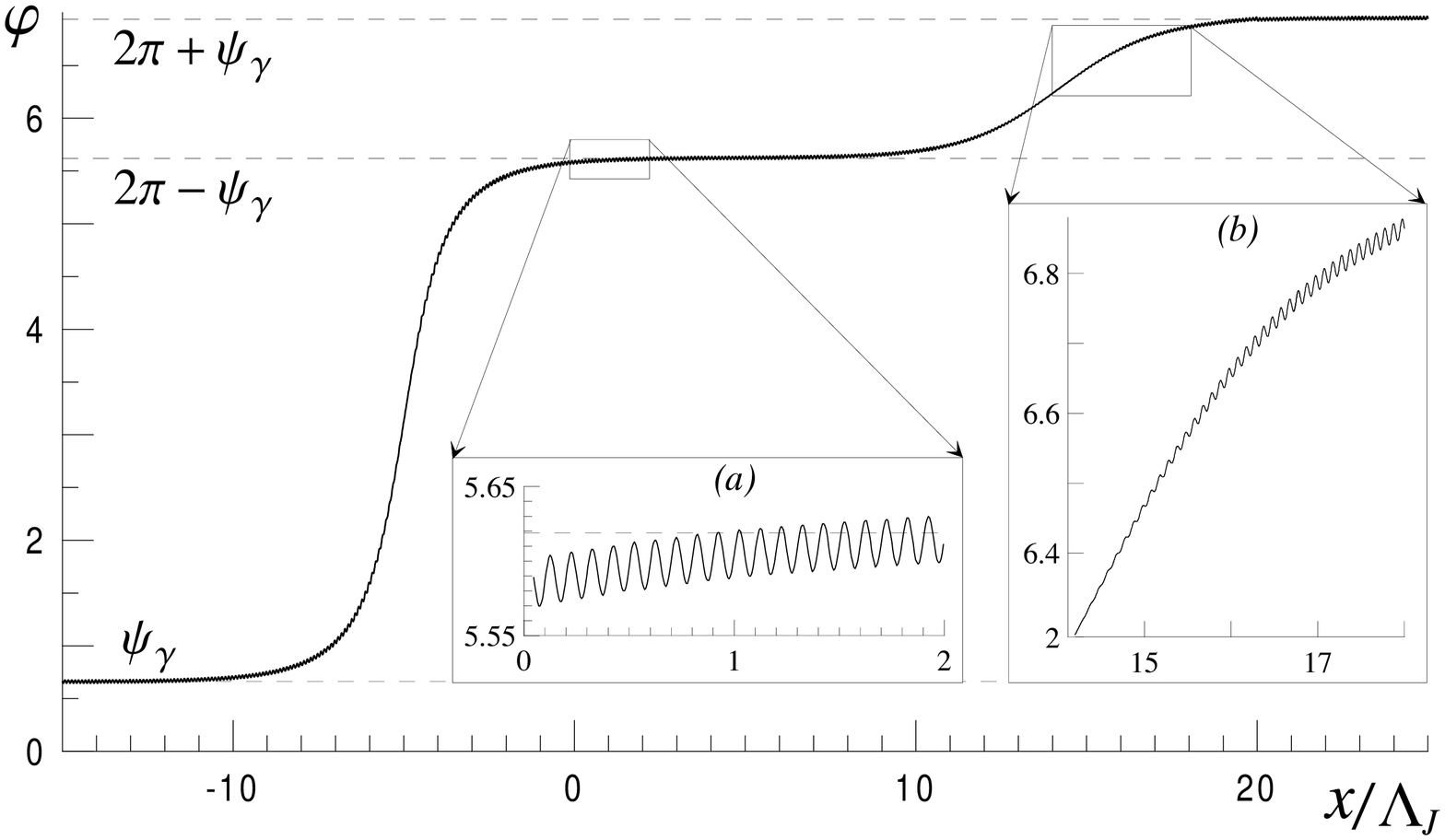} }
\caption{The phase distribution $\varphi(x)$ computed using
$g(x)=g_0\sin (2\pi x/l)$ and $\gamma\approx 1.27$. Two fractional
vortices with $\phi_1\approx 0.21\,\phi_0$ and $\phi_2\approx
0.79\,\phi_0$ are clearly seen, the fine structure of $\varphi(x)$ is
demonstrated in the magnified insets.}
\label{fig_1}
\end{figure}

In Fig.~\ref{fig_1} we show a stable stationary solution for a pair of
fractional vortices. We compute $\varphi(x)$ using the model
 $g(x)=g_0\sin(2\pi x/l)$ with $g_0=100$ and $l =
0.1\Lambda_J$. The value of $\gamma$ calculated by means of
Eq.~(\ref{eq9}) is given by
\begin{equation}
\gamma = {g_0^2\,l^2\over 8\pi^2\Lambda_J^2}.
\label{eq12}
\end{equation}
This yields $\gamma\approx 1.27$ and $\psi_{\gamma}\approx 0.66$; thus
$\phi_1\approx 0.21\,\phi_0$, $\phi_2\approx 0.79\,\phi_0$. As is seen
in Fig.~\ref{fig_1}, the simulation gives the same value of
$\psi_\gamma$. The magnified insets in Fig.~\ref{fig_1} demonstrate
that $\varphi (x)$ indeed consists of a smooth part superimposed with
a small fast oscillating term. Therefore, the numerical simulations
for single fractional vortices confirm the qualitative and
quantitative results of the approximate analytic approach described
above.
\par
\epsfclipon
\par
\begin{figure}
\epsfclipon
\epsfxsize 0.95\hsize
\centerline{ \epsfbox {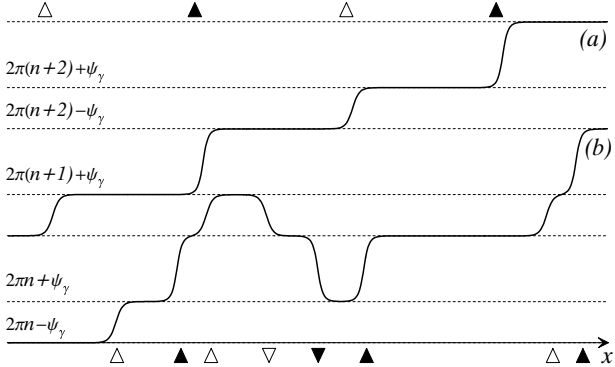}}
\vskip \baselineskip
\caption{Two chains of fractional vortices in a grain boundary with a
periodically alternating critical current density: ({\it a}) an
``ideal" chain, ({\it b}) a chain with vortex-antivortex ``defects''.
Empty triangles mark the positions of the fractional vortices with the
fluxes $\phi_1 <\phi_0/2$, full triangles correspond to
$\phi_2>\phi_0/2$. The up-down orientation of triangles indicate the
field direction of vortices. For this particular calculation we use
$g(x)=g_0\sin (2\pi x/l)$, $g_0=150$, $l=0.1\,\Lambda_J$, which result
in $\gamma\approx 2.85$, $\psi_\gamma\approx 1.21$, and $\phi_1\approx
0.39\phi_0$, $\phi_2\approx 0.61\phi_0$.}
\label{fig_2}
\end{figure}

Consider now a dilute chain of fractional vortices. Let a vortex
with the flux $\phi_1$ be situated somewhere in the chain. The
phase $\psi$ of this vortex changes from $2\pi n -\psi_\gamma$ to
$2\pi n +\psi_\gamma$ with an integer $n$. Therefore, one expects
the phase of an adjacent vortex to start with the value $2\pi n
+\psi_\gamma$ and to end up with $2\pi (n+1) -\psi_\gamma$, the
total phase accumulation of these two vortices being $2\pi$. In
other words, the chain consists of a sequence of pairs of vortices
with fluxes $\phi_1$ and $\phi_2$. This qualitative picture is
confirmed by numerically solving Eq.~(\ref{eq4}). Figure
\ref{fig_2}(a) shows the result of such a calculation for which we
took $g(x)=150\,\sin (20\,\pi x/\Lambda_J)$, that corresponds to
$\gamma\approx 2.85$, $\psi_\gamma\approx 1.21$, and the fluxes
$\phi_1\approx 0.39\phi_0$, $\phi_2\approx 0.61\phi_0$. The final
stationary state of our numerical procedure simulating the
relaxation process, depends on the choice of the initial phase
$\varphi_i(x)$. By taking a proper non-monotonic dependence
$\varphi_i(x)$, we may end up with a stationary solution shown in
Fig.~\ref{fig_2}(b). A remarkable feature of this solution is the
existence of {\it fractional vortex-antivortex pairs} clearly seen
in the simulation of Fig.~\ref{fig_2}(b), the pair with the fluxes
$\pm\phi_1$ is followed by the pair with $\mp\phi_2$.
\par
\par
Next, we study flux patterns for a more realistic case of a grain
boundary with $2N$ facets and a non-periodic alternating critical
current density $j_c(x)$. We treat this case numerically and use a
stepwise $g(x)$ defined as: $g(x)=g_0$ if $a_{i}<x<b_{i}$, and
$g(x)=-g_0$ if $b_{i}<x<a_{i+1}$ ($i=1,\dots,N$). It is convenient to
introduce the random distances $\tilde{a_i}$ and $\tilde{b_i}$ with
$\langle\tilde{a_i}\rangle =\langle\tilde{b_i}\rangle =0$ and a
standard deviation $\sigma_l$ such that
$b_{i}-a_{i}=0.5(l+\tilde{a_i})$ and
$a_{i+1}-b_{i}=0.5(l+\tilde{b_i})$. The simulations start with an
initial phase $\varphi_i(x)$ matching the condition $\varphi_i({\cal
L})-\varphi_i(0) = 2\pi n$. Then the numerical procedure of solving
Eq.~(\ref{eq11}) converges well to a stationary state which depends on
both $\varphi_i(x)$ and $\alpha$ due to the flux pinning induced by
the non-uniformity of the critical current density.
\par
A special role in the description of Josephson boundaries with random
alternating $j_c(x)$ belongs to the stationary state $\varphi_s(x)$
which corresponds to the zero total spontaneous flux and to the
absolute minimum of the Josephson energy ${\cal E}\{\varphi(x)\}$
(defined in a standard way \cite{ABa}). Our simulations show that
$\varphi_s(x)$ is {\it unique} for a given boundary, {\it stable}, and
{\it independent} either of initial guesses $\varphi_i(x)$ or of the
damping constant $\alpha$. Therefore, the phase $\varphi_s(x)$ can
serve as a {\it signature} of each individual boundary. It is
convenient to represent $\varphi_s(x)$ as $\varphi_s(x) =\psi_\gamma
+\xi_s(x)$ with $\psi_\gamma =const$ and the variable part $\xi_s(x)$
having zero average, $\langle\xi_s(x)\rangle =0$, and a typical
amplitude $|\xi_s(x)|<\pi/2$.
\par
\begin{figure}
\epsfxsize 0.95\hsize
\centerline{\epsfbox {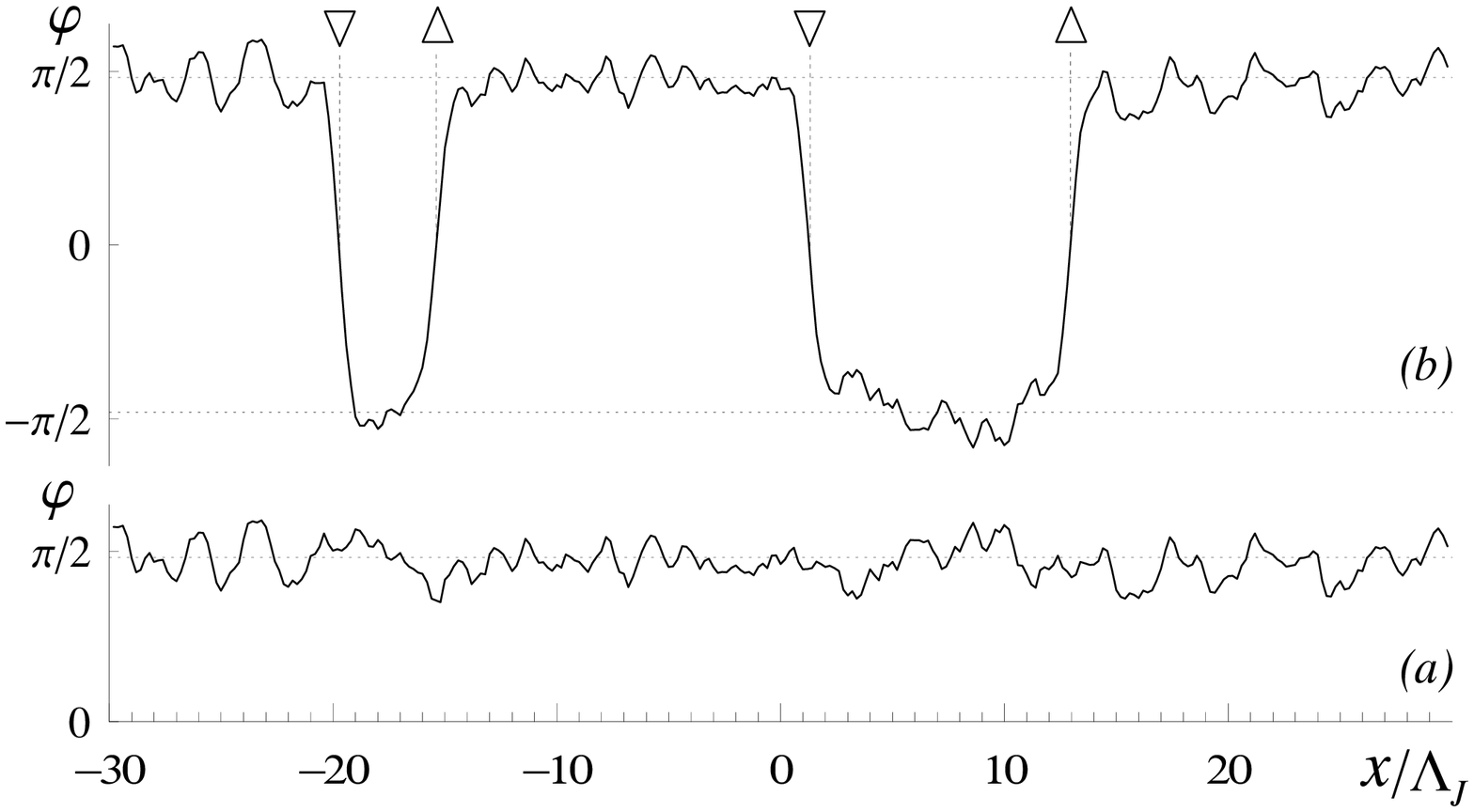}}
\caption{Two stationary solutions $\varphi (x)$ developed in a
zero magnetic field for a stepwise randomly alternating $g(x)$ for two
initial conditions: ({\it a}) preventing, ({\it b}) stimulating
creation of vortices ($g_0=200$, $l=0.1\,\Lambda_J$, $\sigma_l\approx
0.06\,l$, $\psi_{\gamma}\approx 1.515$, $\phi_1\approx 0.48\phi_0$,
$\phi_2\approx 0.52\phi_0$).}
\label{fig_3}
\end{figure}

An example of a computed $\varphi_s(x)$ is shown in
Fig.~\ref{fig_3}(a). For this simulation we took $\varphi_i(x)
=const +\xi_i(x)$ with an arbitrary small $\xi_i(x)$. As stated
above, the resulting phase $\varphi_s(x)$ is independent of
$\varphi_i(x)$. It is worth mentioning that the spontaneous
self-generated flux $\phi_s(x)=\phi_0\xi_s(x)/2\pi$ has a wide
range of length-scales imposed by the random $j_c(x)$. Note also
that randomness of $j_c(x)$ results in a considerably higher
amplitude of the flux variation $\phi_s(x)$ as compared to a
periodic $j_c(x)$; this is seen from comparison of
Fig.~\ref{fig_3}(a) with the insets in Fig.~\ref{fig_2}.
\par
\epsfclipon
\par
It follows from Eq.~(\ref{eq4}) that the stationary solution
$\varphi_s(x)$ generates two series of solutions having the same
Josephson energy as $\varphi_s(x)$: $\varphi^+_n(x)=2\pi
n+\varphi_s(x)$ and $\varphi^-_n(x)=2\pi n-\varphi_s(x)$, where $n$ is
an integer. The average values $\langle\varphi^+_n(x)\rangle=2\pi
n+\psi_\gamma$ and $\langle\varphi^-_n(x)\rangle=2\pi n-\psi_\gamma$
interchange being separated by $\langle\varphi^+_n(x)\rangle -
\langle\varphi^-_n(x)\rangle =2\psi_\gamma$ or by
$\langle\varphi^-_{n+1}(x)\rangle - \langle\varphi^+_n(x)\rangle =2\pi
-2\psi_\gamma$, as is shown in Fig.~{\ref{fig_2}} for a periodic
$j_c(x)$. The gaps between the average values of the stationary phases
$\varphi^+_n(x)$ and $\varphi^-_n(x)$ allow for fractional vortices
with the fluxes $\phi_1=\phi_0\psi_\gamma/\pi$ and
$\phi_2=\phi_0-\phi_1$ as solutions of Eq.~(\ref{eq4}) varying with a
typical length-scale of $\Lambda_J$. An example of a computed $\varphi
(x)$ with two clearly pronounced fractional vortex-antivortex pairs
and small varying part $\xi(x)$ is shown in Fig.~\ref{fig_3}(b).
\par
Consider now a typical flux (phase) pattern for a chain of vortices
for a randomly alternating $j_c(x)$. Assume that the chain starts with
a domain with $\varphi(x)=\varphi_s(x)$, {\it i.e.}, the phase varies
slightly ($|\xi_s(x)|<\pi/2$) around its average value $\psi_\gamma$.
Therefore, $\psi_\gamma$ is the value of $\langle\varphi(x)\rangle$ at
the ``tail'' of the neighboring vortex or antivortex. If the neighbor
carries the flux $\phi_2$, the average phase should increase from
$\psi_\gamma$ to $2\pi -\psi_\gamma$ in the neighbor's domain.
Alternatively the neighbor may carry the flux $-\phi_1$ generating a
decrease of $\langle\varphi(x)\rangle$ from $\psi_\gamma$ to
$-\psi_\gamma$. In general, this flux pattern is similar to the one
arising for a periodic $j_c(x)$. However, the relatively large
amplitudes of the spontaneous flux and the absence of periodicity
within the chain may mask the fractional vortices.
\par
\epsfclipon
\vskip .5\baselineskip
\begin{figure}
\epsfxsize 0.95\hsize
\centerline{\epsfbox {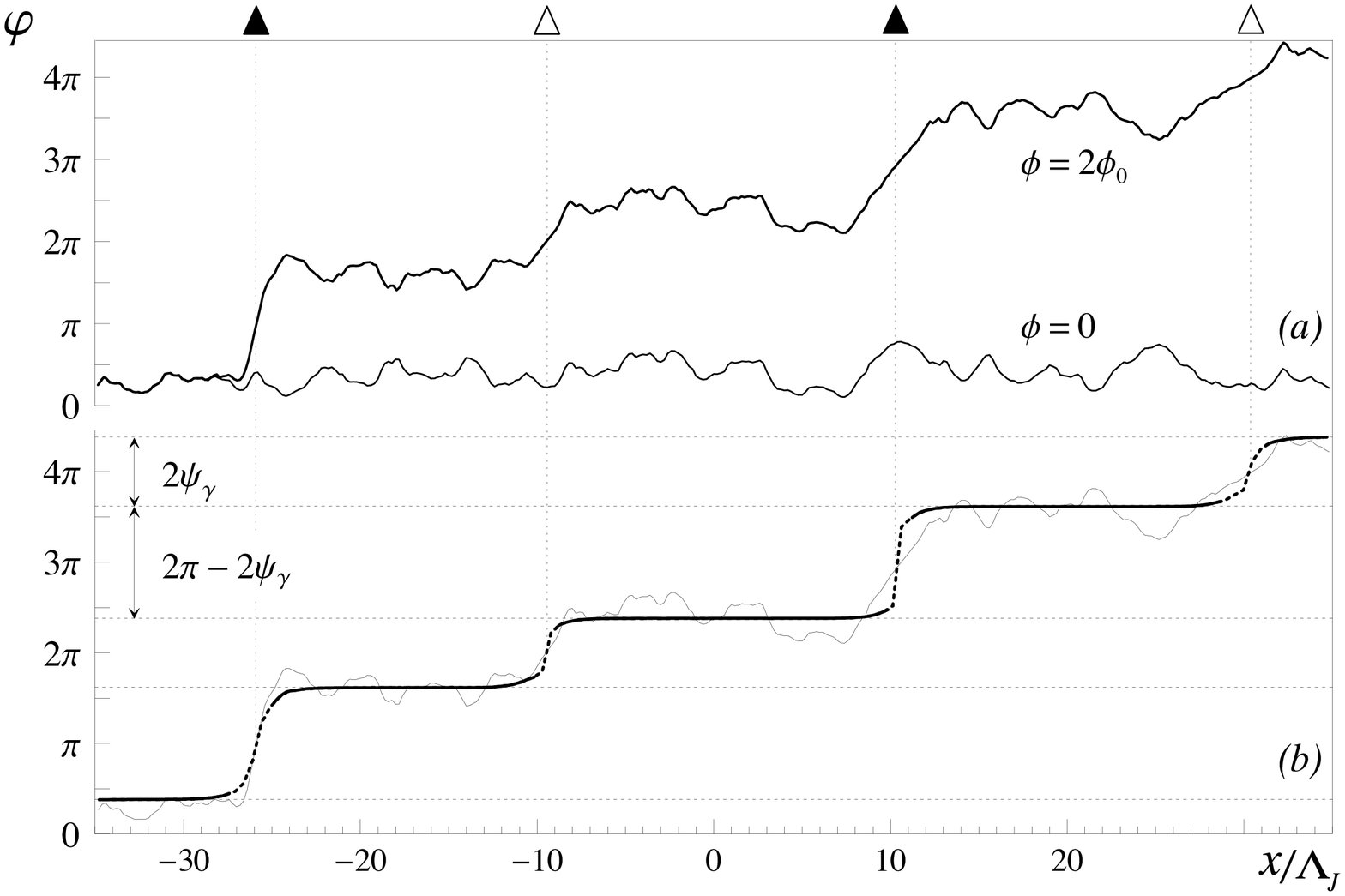}}
\vskip \baselineskip
\caption{($a$) The ``signature" $\varphi_s(x)$ and the
phase $\varphi (x)$ with the total flux $\phi =2\,\phi_0$ at the
boundary calculated for $g_0=90$, $l=0.1\,\Lambda_J$,
$\sigma_l\approx 0.015\,l$, which gives $\psi_\gamma\approx 1.21$,
and $\phi_1\approx 0.39\phi_0$, $\phi_2\approx 0.61\phi_0$. ($b$)
The thin line depicts $\varphi (x)$, the thick line depicts the
phase $\varphi_v(x)$ generated by the fractional vortices and
extracted from the phase $\varphi (x)$.}
\label{fig_4}
\end{figure}
\par
At the top panel of Fig.~\ref{fig_3} we show $\varphi(x)$ obtained
numerically for $g_0=90$, $l=0.1\,\Lambda_J$, and
$\sigma_l=0.015\,l$, the calculated value of $\psi_\gamma$ is
1.21. The bottom curve is the ``signature'' $\varphi_s(x)$
corresponding to the zero total flux, $\phi =0$. The upper curve
is calculated with the same set of input parameters for $\phi
=2\phi_0$. Even a brief examination of the curves shows a striking
correlation between the two: there are domains (e.g.,
$-10<x/\Lambda_J<10$) in which the ``noise'' patterns are nearly
identical, whereas in others (e.g., $10<x/\Lambda_J<30$) the
patterns repeat each other being flipped. This suggests that one
can extract the smooth part of the upper curve $\varphi (x)$ by
properly subtracting the ``signature'' $\varphi_s(x)$.
\par
The subtraction is done as follows. First, we draw the straight lines
$2\pi n\pm\psi_\gamma$ at the graph of $\varphi (x)$, see
Fig.~\ref{fig_4}(b). We see that ``random'' variations of $\varphi
(x)$ are nested on one of these lines everywhere, except a few
relatively sharp jumps from one line to the next; in particular, the
``signature'' $\varphi_s(x)$ is nested at $\psi_\gamma$:
$\varphi_s(x)=\psi_\gamma+\xi_s(x)$. The jumps (or vortices) should be
centered at $\varphi (x)=\pi n$ [where according to Eq.~(\ref{eq4})
$\varphi''(x)=0$]. Then we take a domain situated between lines $\pi n$
and $\pi (n+1)$ and form $\varphi_v(x)=\varphi (x)\mp\xi_s(x)$,
choosing the minuses if the random parts of $\varphi (x)$ and
$\varphi_s(x)$ are identical, and the plus otherwise. The curve
$\varphi_v(x)$ shown by a thick curve at Fig.~\ref{fig_4}(b) is
remarkably {\it smooth}; clearly it represents the fractional vortices
described above within the model with periodic $j_c(x)$.
\par
In conclusion, we have shown by numerical simulations that two types
of fractional vortices with fluxes $\phi_1<\phi_0/2$ and
$\phi_2=\phi_0-\phi_1>\pi/2$ exist at 45$^{\circ}$ [001]-tilt grain
boundaries in YBa$_2$Cu$_3$O$_{7-x}$ films exhibiting spontaneous flux
in zero field cooled samples. Faceted grain boundaries are treated as
Josephson junctions with alternating critical current density. We show
how to extract fractional vortices from the data on spontaneous flux
distribution.
\par
One of us (RGM) is grateful to J.R. Clem, V.G. Kogan, and J. Mannhart
for useful and stimulating discussions. This research is supported in
part by grant No.~96-00048 from the United States -- Israel Binational
Science Foundation (BSF), Jerusalem, Israel.
\par
\end{multicols}
\end {document}